# Tuning the Surface States of $Fe_3O_4$ Nanoparticles for Enhanced Magnetic Anisotropy and Induction Efficacy

Kyle A. Portwin\*, Pablo Galaviz, Xiaoning Li, Chongyan Hao, Lachlan A. Smillie, Mengyun You, Caleb Stamper, Richard Mole, Dehong Yu, Kirrily C. Rule, David L. Cortie & Zhenxiang Cheng†

\* kyleportwin@gmail.com † cheng@uow.edu.au

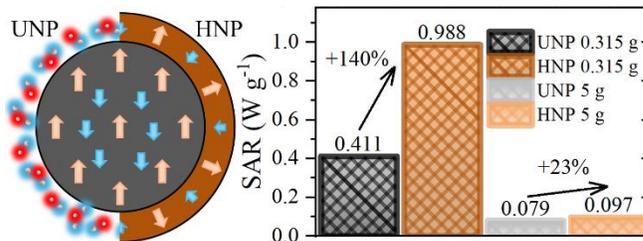

**ABSTRACT:** Magnetite ($Fe_3O_4$) nanoparticles are crucial for biomedical applications, including magnetic hyperthermia, targeted drug delivery, and MRI contrast enhancement, due to their biocompatibility and unique physicochemical properties. Here, we investigate how surface states influence their induction performance. Heat treatment removes surface water and FeOOH, forming a γ-$Fe_2O_3$ shell, as confirmed by synchrotron powder diffraction, neutron powder diffraction, thermogravimetric analysis, X-ray photoelectron spectroscopy, X-ray absorption spectroscopy, and time-of-flight inelastic neutron spectroscopy. AC magnetic susceptibility measurements reveal that this surface modification enhances magnetic anisotropy and reduces the spin relaxation time, leading to a 140% increase in the specific absorption rate. Additionally, the increased anisotropy suppresses the low-temperature clustered spin-glass transition and raises the blocking temperature. These findings highlight surface-state engineering as a powerful approach to optimizing $Fe_3O_4$ nanoparticles for biomedical applications.

Iron oxide nanoparticles, particularly magnetite ($Fe_3O_4$), have attracted significant interest in biomedical applications due to their unique physicochemical properties, including excellent biocompatibility, superparamagnetic behavior, and efficient heat generation under alternating magnetic fields.[1-7] These characteristics make $Fe_3O_4$ nanoparticles promising candidates for magnetic hyperthermia therapy, targeted drug delivery, and MRI contrast enhancement.[3-5]

The induction efficacy of magnetic nanoparticles is closely linked to their specific absorption rate (SAR), which quantifies the power absorbed per unit mass under an AC magnetic field. The SAR is primarily governed by the nanoparticles' magnetic properties and surface characteristics.[8-13] In particular, surface functionalization plays a critical role in determining induction performance. Notably, adsorbed water can significantly alter the magnetic properties and heating efficiency of these nanoparticles.[14] Despite the ubiquity of surface water in nanoparticles, its precise impact on induction efficacy and the overall therapeutic performance of $Fe_3O_4$ nanoparticles in hyperthermia remains poorly understood. A deeper understanding of the role of surface groups, such as adsorbed water molecules, in modulating magnetic anisotropy, spin relaxation times and SAR is essential for the rational design of $Fe_3O_4$ nanoparticles optimized for biomedical applications.

In this study, we systematically investigate the influence of surface water on the induction capabilities and magnetostatic interactions of $Fe_3O_4$. Our findings reveal that surface water and hydroxyl groups (such as FeOOH) impede induction, leading to a substantial reduction in SAR. Through analysis of the complex AC magnetic susceptibility ($\chi''$) we show that untreated nanoparticles (UNP), with surface water and FeOOH, exhibit a clustered spin-glass transition, which is suppressed in heat-treated nanoparticles (HNP). Additionally, by examining the superparamagnetic transition, we demonstrate that removing the surface adsorbates significantly enhances magnetic anisotropy (K) and the relaxation time ($\tau_0$). These results underscore the critical role of surface states in governing the magnetic properties of $Fe_3O_4$ nanoparticles, offering key insights into their optimization for induction-based biomedical applications. This work advances fundamental understandings of the interplay between surface chemistry and magnetic properties in $Fe_3O_4$, providing a framework for the design of high-performance nanoparticles for induction applications.

RESULTS AND DISCUSSION

**Structure, chemical valences and bonding.** A comparison of UNP and HNP synchrotron powder diffraction phases is presented in Fig. 1A. The crystal structures were refined using the model from [15] (Fig. 1B). $Fe_3O_4$ crystallizes in the cubic F d - 3 m (No. 227) space group, with an inverse spinel structure, where $Fe^{+3}$ ions occupy both tetrahedral and octahedral sites and $Fe^{2+}$ ions are confined to octahedral sites.[16] Lattice parameters for UNP and HNP were determined to be (8.3588±0.0001) Å and (8.3457±0.0001) Å, respectively, within ~0.013 Å of each other. These values are closer to the lattice parameters of maghemite (γ-$Fe_2O_3$, with a = 8.345 Å) compared to the expected value for pure $Fe_3O_4$ (a = 8.395 Å).[15, 17-20] Since synchrotron powder diffraction is a surface-sensitive technique, and with the increased surface-to-volume ratio of nanoparticles, the obtained lattice parameters represent a surface oxidation with a strong signal from γ-$Fe_2O_3$. The decrease in lattice parameters of HNP suggests a greater percentage of γ-$Fe_2O_3$ on the surface compared to UNP.[20] Scherrer analysis of the full-width-at-half-maximum (FWHM) yielded nanoparticle diameters of (16.3±0.2) nm (UNP) and (16.5±0.3) nm (HNP), indicating that heat treatment did not significantly alter particle size. The goodness-of-fit ($\chi^2$) values are 1.81 (UNP) and 1.89 (HNP), with a more negative microstrain after heat treatment (from -1397.7 to -4009.4), suggesting an increase in local strain.

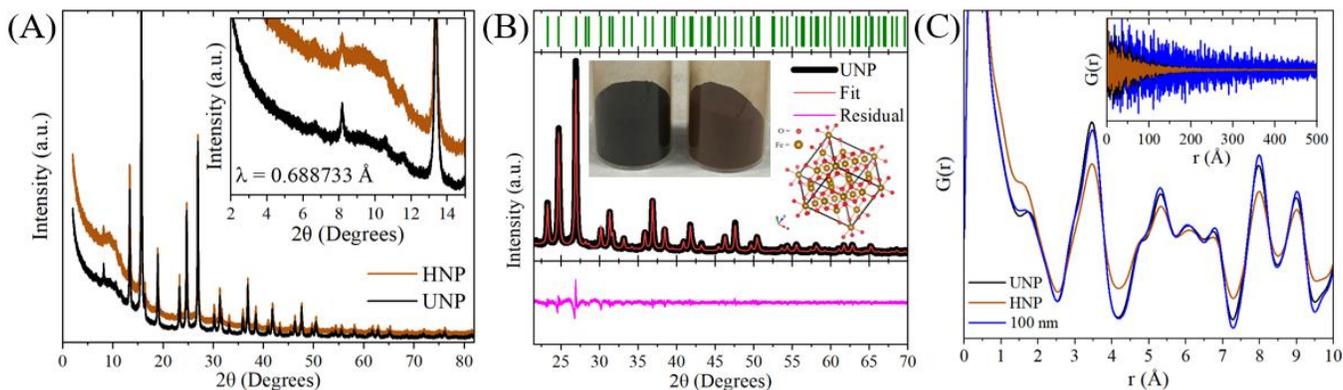

Figure 1. (A) Synchrotron powder diffraction spectra for UNP and HNP, the inset highlights the amorphous peak. (B) Rietveld refinement of UNP with planes of reflection above, and a residual of fit subtracted from experimental data below. The insets show the samples; with UNP on the left and HNP on the right, and the unit cell of $Fe_3O_4$. Note that UNP are black, while HNP are a reddish-brown, corresponding to the colors of $Fe_3O_4$ and $\gamma\text{-}Fe_2O_3$, respectively. (C) Pair distribution function at low bond lengths, the inset shows high bond lengths.

A difference between UNP and HNP is the broad peak at ~10° in the inset of Fig. 1A, attributed to increased diffuse scattering from an amorphous layer.[21-23] This glassy layer introduces localized strain gradients, as evidenced by the increase in negative microstrain shift and lattice parameter contraction. To further probe local structural changes, we analyzed the pair distribution function (PDF) (Fig. 1C), which provides atomic pair probabilities $G(r)$ as a function of separation distance $r$.[24, 25] The PDF confirms that peak positions remain unchanged except for a low-$r$ peak at ~1.6–1.7 Å, which shifts to lower $r$ and increases in intensity for HNP. While UNP closely resemble bulk 200 nm $Fe_3O_4$ particles, HNP exhibits broader, less intense peaks, suggesting greater atomic disorder at the surface. At long range, signals for both UNP and HNP decay around ~160 Å, corresponding to the finite nanoparticle size. In contrast, the 200 nm reference sample maintains atomic correlations at much larger $r$, as finite size is not a limiting factor. These results show that heat treatment does not alter internal bond lengths or particle size, however, it induces surface disorder and unit cell contraction.[26] We hypothesise that this is due to $\gamma\text{-}Fe_2O_3$ formation from FeOOH dehydroxylation, and that the PDF peak shift represents $\gamma\text{-}Fe_2O_3$ having shorter Fe–O bonds than FeOOH.[27-31]

To test this hypothesis and investigate the origin of strain and amorphization, we perform thermogravimetric analysis (TGA), Fig. 2A and 2B. This reveals four key stages during the heat-treatment process, labelled i), ii), iii) and iv), which closely match previous reports on the thermal behavior of $Fe_3O_4$ with adsorbed FeOOH and $H_2O$ under inert conditions:[22, 32-41]

(i) (Room temperature – 150 °C) UNP exhibit a mass loss of ~1.9%, corresponding to the desorption of $H_2O$.

(ii) (150 °C – 250 °C) A subsequent mass increase/plateau reflects competing processes: FeOOH → $\gamma\text{-}Fe_2O_3$ + $H_2O$, where dehydroxylation-induced mass loss is offset by density-driven mass gain.

(iii) (250 °C – 350 °C) A continued loss of ~1.1% from hydroxide and $H_2O$ loss in the dehydroxylation process.

(iv) (350 °C – 400 °C + isothermal) During annealing, a steady mass loss of approximately 0.2% is observed. This gradual decrease reflects the transformation of magnetite ($Fe_3O_4$, density 5.17–5.18 g cm$^{-3}$)[42] to maghemite ($\gamma\text{-}Fe_2O_3$, density 4.87–4.90 g cm$^{-3}$)[43]. In the absence of external oxidants, due to the inert argon atmosphere, this phase change is driven by the redistribution of oxygen within the crystal lattice.

It is worth noting that $\gamma\text{-}Fe_2O_3 \rightarrow \alpha\text{-}Fe_2O_3$ (hematite) is not observed below 400 °C.[40, 44]

X-ray photoelectron spectroscopy (XPS) with etching was used to probe the core/shell composition and electronic states of the nanoparticles. In the Fe L edge spectrum (Fig. 2C) there exist two main peaks corresponding to $2p_{3/2}$ and $2p_{1/2}$ electrons and a shake-up satellite for each of these peaks. Table 1. compares each of the peak positions of UNP and HNP before and after etching, effectively the shell and core of the particles respectively, in comparison to literature values.

Table 1. XPS Fe L edge peak positions.

| | $Fe2p_{3/2}$ (eV) | Satellite (eV) | $Fe2p_{1/2}$ (eV) | Satellite (eV) |
|---|---|---|---|---|
| UNP | 710.8 | 718.8 | 724.4 | 733.4 |
| UNP Etch | 710.3 | N/A | 723.4 | N/A |
| HNP | 710.7 | 719 | 724.4 | 733.4 |
| HNP Etch | 710.4 | N/A | 723.6 | N/A |
| $Fe_3O_4$ [45] | 710.5 | N/A | 724 | N/A |
| $\gamma\text{-}Fe_2O_3$ [45] | 711 | 718.8 | 724.6 | 733.5 |

The peak positions before etching remain fairly constant, and closer to the values of $\gamma\text{-}Fe_2O_3$. After etching, we observe a down-shift in the $2p_{3/2}$ and $2p_{1/2}$ binding energies to values more closely aligned with $Fe_3O_4$. Another key difference is the vanishing of satellite peaks, characteristic to $\gamma\text{-}Fe_2O_3$, upon etching, which is an excellent indication of the $Fe_3O_4/\gamma\text{-}Fe_2O_3$ core/shell structure.[45-51] T. Fujii et. al.[51] directly show these corresponding peak positions and shifts for $Fe_3O_4$ and $\gamma\text{-}Fe_2O_3$ confirming our results. Furthermore, S. Ali et. al.[52] present similar results for etching $Fe_3O_4$ with a $Fe_3O_4/\gamma\text{-}Fe_2O_3$ core/shell structure.

In the O K edge spectrum (Fig. 2D), the expected main peak position for $Fe_3O_4$ is 529.7 eV, while $\gamma\text{-}Fe_2O_3$ is 530.0 eV. The slight shift to higher energies in HNP reflects the increase $\gamma\text{-}Fe_2O_3$ compared to the UNP.[45, 47-49, 53] The main peak position of etched nanoparticles more closely resembles $Fe_3O_4$.

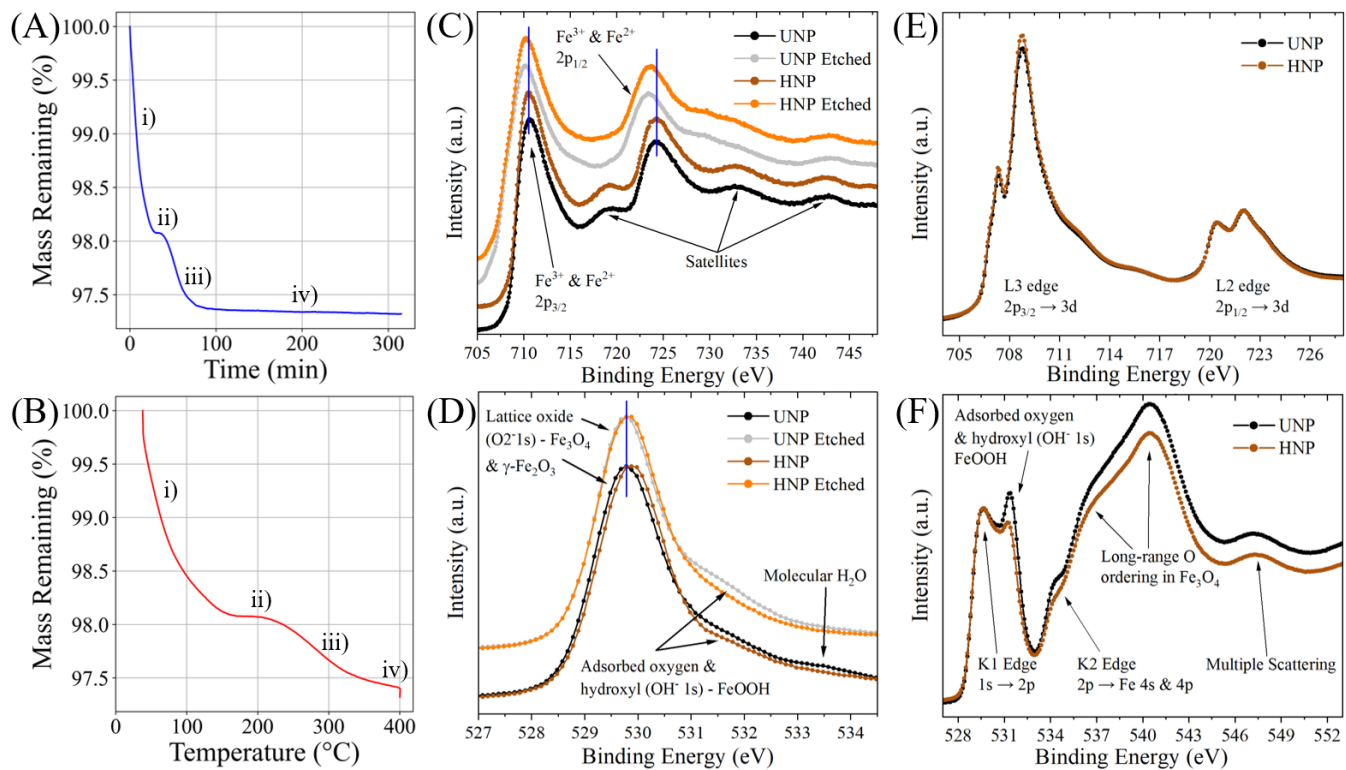

Figure 2. (A) and (B) TGA curves, displaying the mass lost with respect to time and temperature, respectively. Key stages are labelled. (C) and (D) XPS results for the Fe L edge and the O K edge, respectively, labels indicate peaks, and blue lines highlight peak shifting. Spectra are normalised to the highest peak position and shifted in intensity to highlight differences. (E) and (F) XAS results for the Fe L edge and O K edge, respectively, labels indicate peaks. Error bars are contained within the data points.

The shoulder peak at 531–532.5 eV, associated with adsorbed hydroxyl groups and $H_2O$, diminishes in intensity for HNP and etched HNP, supporting FeOOH dehydration. [45, 47-49, 53] There is also a distinct surface water peak in UNP which diminishes in HNP and etched samples, consistent with water removal during annealing and etching. XPS peak deconvolution reveals that after heat treatment, oxygen content decreases from 64.65% to 62.23%, while $Fe^{3+}$ and $Fe^{2+}$ percentages increase from 26.19% to 27.77% and 9.16% to 10.00%, respectively. This is consistent with the dehydration of FeOOH, following FeOOH → $\gamma$-$Fe_2O_3$ + $H_2O$, and the formation of a $\gamma$-$Fe_2O_3$ rich surface. The concurrent increase in both $Fe^{2+}$ and $Fe^{3+}$ suggests that, in addition to shell formation, $Fe_3O_4$ beneath the removed FeOOH becomes more exposed. Thus, the observed surface chemistry may reflect a partial oxidation coupled with core exposure, rather than complete transformation to $\gamma$-$Fe_2O_3$.

X-ray absorption spectroscopy (XAS), another surface-sensitive technique, further elucidates the electronic structure and oxidation states.[54] The Fe L edge spectra (Fig. 2E) displays four characteristic peaks at 707.3 eV, 708.8 eV, 720.4 eV, and 722 eV, corresponding to $Fe^{2+}$ and $Fe^{3+}$ in different coordination environments. While FeOOH, $\gamma$-$Fe_2O_3$ and $Fe_3O_4$ have very similar Fe L edge spectra, the 4 sharp peaks more closely align with FeOOH and $\gamma$-$Fe_2O_3$ spectra in the literature, compared to $Fe_3O_4$ which displays two much broader peaks.[55-61] The intensity increases post heat treatment, indicating an enrichment of surface Fe, while peak positions remain largely unchanged, reflecting the persistence of surface $\gamma$-$Fe_2O_3$. The O K edge spectra (Fig. 2F) provides additional insight into surface transformations. The K1 edge appears closer to the shape of $\gamma$-$Fe_2O_3$ (two distinct peaks) compared to $Fe_3O_4$ (one broader peak).[52, 56-60, 62-64] The first peak at 529.6 eV remains consistent post heat treatment, meanwhile, the 531.3 eV peak, shifts slightly and decreases in intensity. We link this decrease in intensity to the dehydration of FeOOH (which shows a much larger intensity in this second peak) to $\gamma$-$Fe_2O_3$. S. Chen *et. al.*[60] confirm this by presenting the O K edge of FeOOH when heated by the XAS ion gun which directly shows the dehydration of FeOOH to $Fe_2O_3$. The higher energy end of the spectrum closely resembles a combination of $\gamma$-$Fe_2O_3$ and $Fe_3O_4$, which are extremely similar in shape. The decrease in intensity of higher energy peaks can be attributed to the loss of FeOOH to $\gamma$-$Fe_2O_3$.[52, 56-60, 62-64]

**Electron Microscopy Characterization.** Transmission electron microscopy (TEM) images (Fig. 3A-D) confirm that UNP and HNP exhibit comparable crystallinity and morphology. Particle size distributions (Fig. 3E and 3F) yield average diameters of (16.2±0.2) nm for UNP and (16.0±0.2) nm for HNP, consistent with supplier specifications. While these values are within experimental uncertainty, they slightly differ from Rietveld refinement results. However, the absence of size change aligns with PDF analysis, indicating that heat treatment does not alter core dimensions. Most nanoparticles are spherical, though some exhibit cubic and elliptical morphologies.

Scanning electron microscopy (SEM) combined with energy-dispersive spectroscopy (EDS) mapping (Fig. 3G and 3H) confirms the homogeneous distribution of Fe and O in both samples. Fig. 3I displays the colour contour EDS maps for HNP confirming the presence of Fe and O together on a C grid.

The EDS spectra reveal key Fe peaks at ~0.8 keV, ~6.4 keV, and ~7.1 keV, and an O peak at ~0.5 keV. [65-68] Quantitative analysis shows a decrease in oxygen content from 40.23% to 30.61% after heat treatment, with a corresponding increase in

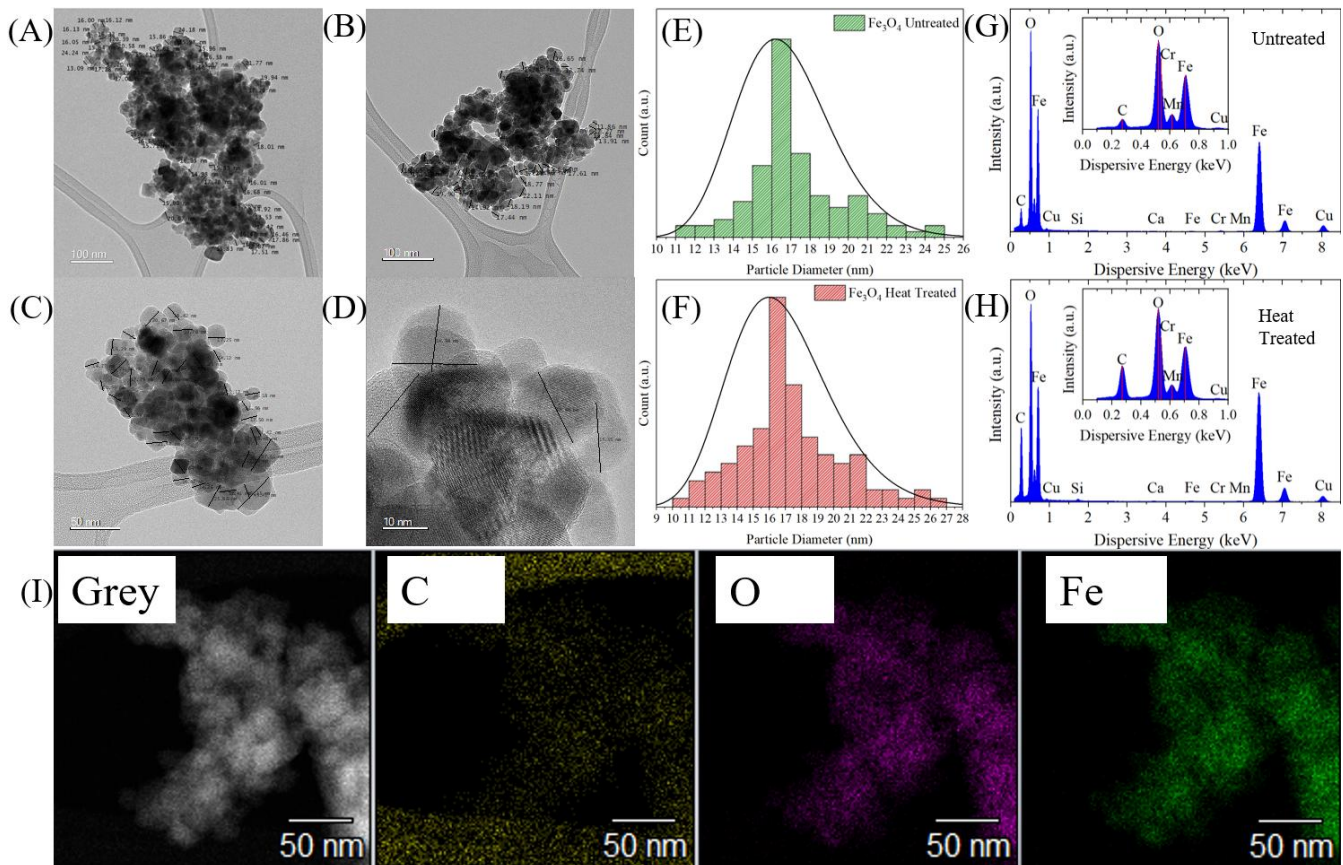

Figure 3. (A), (B), (C) and (D) show different scaled images of nanoparticles at various positions on the carbon grid. (E) and (F) present the particle size distribution histograms with logarithmic fits of UNP and HNP, respectively. (G) and (H) show the EDS spectra for UNP and HNP, respectively. The insets zoom into the low energy 0-1 keV region to reveal convoluted peaks. (I) shows the EDS maps for HNP.

Fe content (59.77% → 69.39%), supporting the removal of surface water and hydroxyl species. Additionally, the Fe 0.8 keV / 6.4 keV peak intensity ratio changes post-treatment, suggesting a modified ratio of FeOOH, $\gamma$-$Fe_2O_3$ and $Fe_3O_4$ in the samples.

**Time-of-flight neutron scattering.** To investigate the presence of surface water and hydroxyl groups, we performed time-of-flight inelastic neutron scattering (TOF-INS) to analyze the structure and quasielastic neutron scattering (QENS) signal in both $Fe_3O_4$ nanoparticle samples. Since neutrons are highly penetrating, they provide insight to the entire nanoparticle core and shell structure. Fig 4A-D shows the scattering function, $S(Q,\omega)$, for UNP and HNP at 300 K and 200 K. Above the $S(Q,\omega)$, we show the elastic signal $S(Q)$, effectively neutron powder diffraction. The powder diffraction spectra for 200 K UNP and HNP have been Rietveld refined (Supplementary Information Fig S1) which reveal a high concentration, 87.6(4)% to 12.4(4)%, of $Fe_3O_4$ to $\gamma$-$Fe_2O_3$, respectively. The weight fraction of $\gamma$-$Fe_2O_3$ increases in HNP to 13.8%, consistent with the dehydration of FeOOH to form more $\gamma$-$Fe_2O_3$. We also see the lattice parameter of $Fe_3O_4$ and $\gamma$-$Fe_2O_3$ shift to lower values (~0.02 Å) in agreement with the unit cell contraction observed in synchrotron powder diffraction refinements.

At 300 K (Fig. 4A and 4B) UNP exhibit a strong QENS signal, characteristic of dynamic water molecules [69-71], see Supplementary Information Figure S2 for the water QENS. Contributions from diffusive hydrogen in FeOOH appear as weak signals superimposed on the broader QENS component, consistent with dynamics reported by E. Brok et. al.. [72,73] In contrast, the QENS signal is significantly reduced in HNP, confirming the removal of water upon heat treatment. [69] The remaining spectral weight is attributed to hydrogen in FeOOH, though diminished in intensity compared to UNP, suggesting that not all FeOOH has been fully converted to $\gamma$-$Fe_2O_3$.

Fig. 4C and 4D display $S(Q,\omega)$ at 200 K, where both samples show a significantly weaker QENS signal due to the reduced mobility of surface water and FeOOH molecules, which freeze at this temperature. [74] The suppression of translational diffusion is further evident in Figure 4E, which presents $S(\omega)$ on a logarithmic scale. At 300 K, UNP exhibit a broad, intense QENS signal indicative of liquid-like diffusion, transitioning to localized motion at shorter length scales. In contrast HNP at 300 K primarily show rotational or confined diffusion. Analysis of the elastic incoherent structure factor (EISF) and FWHM (Supplementary Information QENS Analysis Figure S3) provides additional insights, revealing long-range translational diffusion in UNP at 300 K for larger length scales, while all other conditions exhibit localized dynamics.

Fig. 4F and 4G compare the phonon density of states (DOS) for both samples alongside data for water and ice. At 300 K, the DOS of UNP strongly overlaps with that of liquid water, confirming the dominant role of water dynamics in the observed signal. Upon heat treatment, the DOS intensity decreases and shifts, reflecting reduced water contributions. At 200 K (Fig. 4G), the DOS of UNP exhibits minor features associated with ice, corresponding to the freezing of surface water. The DOS of HNP, however, shows minimal overlap with either liquid water or ice, consistent with water removal.

Fig. 4H focuses on the low-energy acoustic regime of the DOS. A hallmark of liquid dynamics is a linear DOS at low energy, which arises from overdamping of normal modes. [75, 76] At 300 K, UNP exhibit a clear linear DOS, whereas HNP follows the quadratic form predicted by the Debye model, indicative of solid-like behavior. At low temperatures, both UNP and HNP, as well as ice, exhibit a quadratic, solid-like DOS, as expected.

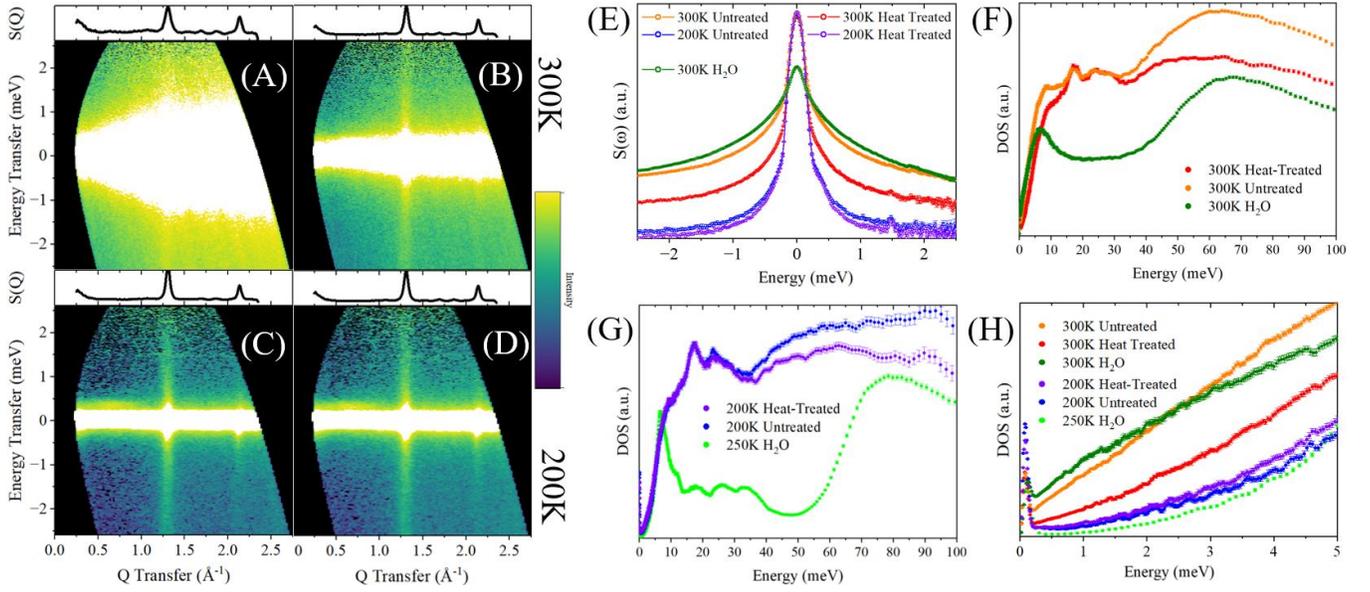

Figure 4. Time-of-flight inelastic neutron scattering results. (A) and (C), UNP S(Q,ω) at 300 K and 200 K, respectively. (B) and (D), HNP S(Q,ω) at 300 K and 200 K, respectively. (E) S(ω) on a log scale to display QENS signal. (F) and (G) are the DOS of nanoparticle samples and water at room temperature and low temperature, respectively. (H) DOS of (F) and (G) in the low energy acoustic region.

The results of these combined characterization techniques confirm UNP consist of an $Fe_3O_4$ core with surface FeOOH and adsorbed water, while HNP develop a strained $\gamma$-$Fe_2O_3$ shell upon heat treatment. Despite this surface transformation, core size and morphology remain relatively unchanged, consistent with previous reports on $Fe_3O_4$ nanoparticle dehydration. [12, 13]

**Magnetic properties and magnetic induction efficacy.** Fig. 5A shows the temperature evolution of the nanoparticles under magnetic induction heating, with the inset displaying the sample inside the coil. The 0.315 g samples reach saturation at temperatures of ~323 K and ~348 K for UNP and HNP, respectively, as indicated by the plateau of the heating curves. The 5 g samples achieve higher temperatures due to a larger total surface area for heat generation and magnetic interactions. However, heating above 373 K risks altering surface water content, so the experiment was terminated at this point. The induction heating efficiency is quantified by the SAR, given by:

$$SAR = \frac{C \cdot \Delta T}{m \cdot \Delta t} \quad (1)$$

where ΔT (K) is the temperature change over the time interval Δt (s), m is the nanoparticle mass and C ~ 0.65 J $g^{-1}$ $K^{-1}$ [77, 78] is the specific heat capacity. The SAR is determined from the initial, most linear region of the heating curve, before thermal equilibrium effects slow the temperature rise. For the first 20 s of each curve in Fig. 5A, the ΔT/Δt has been determined, and SARs have been calculated using equation (1), results are presented in Table 2. The SAR increases by 140% for 0.315 g and a 23% for 5 g samples, highlighting enhanced efficiency in HNP, particularly at lower sample masses relevant to induction applications.

Table 2. SAR determined by the induction rate of differing masses of UNP and HNP $Fe_3O_4$.

| Sample | ΔT/Δt (K $s^{-1}$) | SAR (W $g^{-1}$) |
|---|---|---|
| 0.315 g UNP | 0.199±0.008 | 0.411±0.017 |
| 0.315 g HNP | 0.479±0.009 | 0.988±0.019 |
| 5 g UNP | 0.608±0.015 | 0.079±0.002 |
| 5 g HNP | 0.745±0.010 | 0.097±0.001 |

To understand the origin of this SAR enhancement, we investigate the DC and AC magnetic properties. Fig. 5B and 5C show magnetic hysteresis loops from 5 K to 350 K, for UNP and HNP, respectively, with the inset of both images displaying a zoomed in perspective of the hysteresis at 350 K, 250 K and 5 K measurements. Figure 5D compares the temperature dependent magnetic coercivity and saturation magnetization of both samples. At 300 K, the coercivity is nearly zero in both samples, as expected behaviour for superparamagnetic nanoparticles, and increases with decreasing temperature as more particles are blocked from spontaneous spin flip renormalization (flipping). At 350 K, most nanoparticles are in the superparamagnetic state, though a fraction of larger sized particles remain blocked. Notably, HNP exhibit lower coercivity at all temperatures, indicating a greater fraction of the sample is superparamagnetic.

The saturation magnetization is lower for HNP at all temperatures, consistent with the formation of $\gamma$-$Fe_2O_3$, which has lower magnetization than $Fe_3O_4$. The saturation magnetization, which ranges from ~54–66 emu $g^{-1}$, is consistent with other nanoparticle $Fe_3O_4$ values in the literature,[21, 32, 34-38, 79, 80], and is notably lower than the value for bulk $Fe_3O_4$.

The superparamagnetic nature of nanoparticles is dependent on particle size, shape and dipolar interactions. To assess the superparamagnetic blocking temperature, Fig. 5E shows the zero-field-cooled (ZFC) and field-cooled (FC) curves for both samples, under a 100 Oe applied field. Due to the particle size distribution, the blocking temperature is not immediately evident but can be taken to the left of the ZFC maximum,[81] closer to where the FC curves begin to decrease (see inset of Fig 5E). This places the blocking temperature at ~200–225 K.

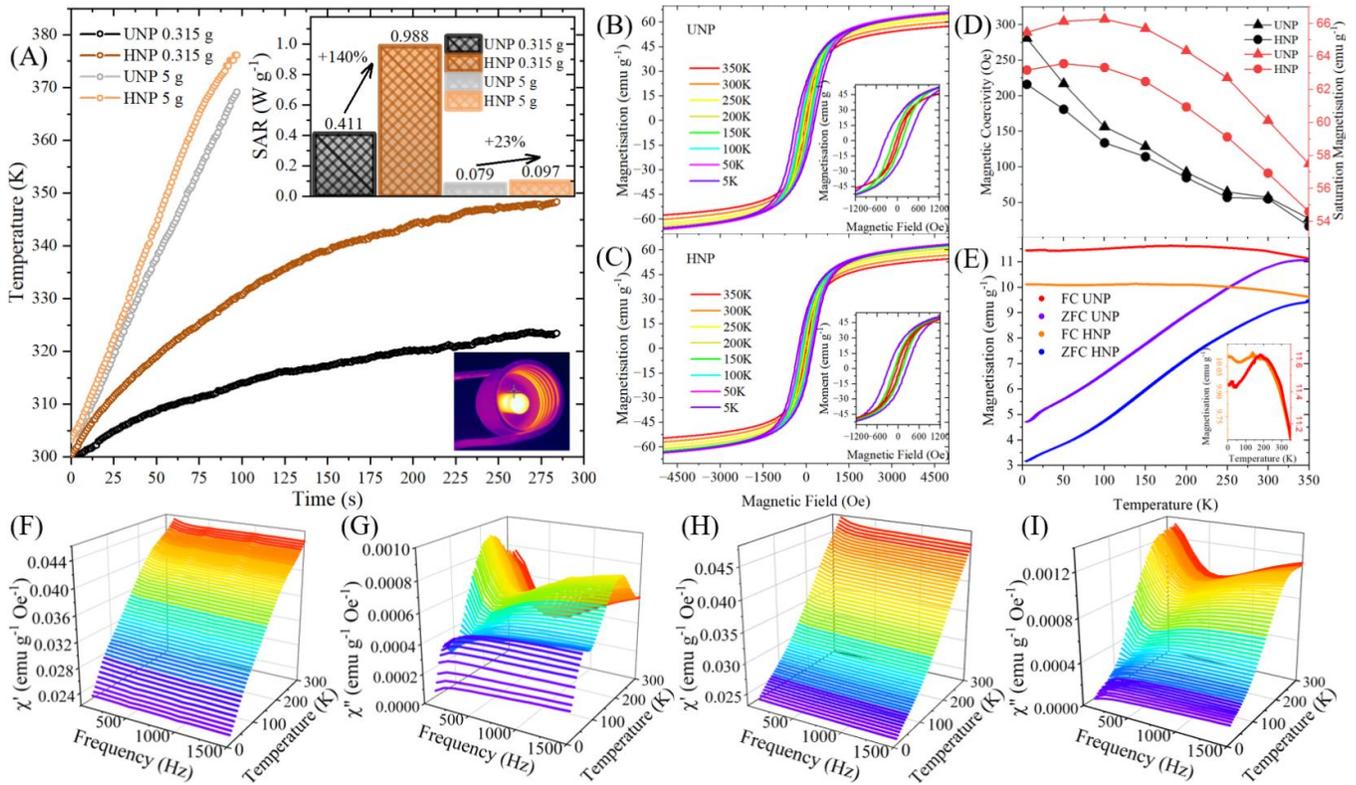

Figure 5. Magnetic characterization. (A) Induction results for 0.315 g and 5 g of UNP and HNP. The insets shows the experimental setup through the thermal imaging camera, and the calculated SAR values. (B) and (C) show temperature dependent hysteresis loops for UNP and HNP, respectively. Each inset zooms in on the middle of the loops to accentuate the coercivity. (D) Temperature dependence of the magnetic coercivity and the saturation magnetisation. (E) ZFC and FC magnetization curves, with the inset highlighting the differences between the FC curves. (F) and (G) show the frequency and temperature dependence of $\chi'$ and $\chi''$ in UNP and (H) and (I) show the frequency and temperature dependence of $\chi'$ and $\chi''$ in HNP. Error bars are contained within data points.

Fig. 5F and 5G show the real ($\chi'$) and imaginary ($\chi''$) components of the AC susceptibility for UNP, while Fig. 5H and 5I present $\chi'$ and $\chi''$ for HNP. The in-phase component, $\chi'$, reflects how well the nanoparticle spin-flips track the AC field. HNP exhibits increased $\chi'$ across the whole temperature range, suggesting improved dynamic response to the AC field. A sharp decrease in $\chi'$ occurs at the blocking temperature, marking the transition from superparamagnetism to ferromagnetism.

The out-of-phase component, $\chi''$, represents power dissipation and shows two temperature-dependent peaks in UNP (~50 K and ~200 K), while HNP exhibits a single peak at ~ 225 K. These features align with the ZFC-FC results. The magnitude of $\chi''$ is generally higher in HNP, consistent with an increased induction rate/SAR. The temperature dependence of $\chi''$ reveals distinct magnetic relaxation processes in UNP and HNP. The freezing temperatures ($T_f$) of these transitions exhibit frequency-dependent shifts (Figure 6A and 6B).[11, 82-86] To determine the nature of these transitions, an external DC field was applied (Supplementary Information Figure S4) which shows the low-temperature peak shifts to lower values and the high-temperature peak shifts to significantly higher values, indicative of a spin-glass transition and superparamagnetic transition, respectively. The absence of a spin-glass transition in HNP implies that surface states play a key role in governing the low-temperature spin dynamics.

To further analyse these dynamics, we apply the Vogel-Fulcher Law, which describes collective spin dynamics in disordered magnetic systems:

$$\tau = \tau_0 \exp\left(\frac{Kv}{k_B(T_f - T_0)}\right) \quad (2)$$

where $\tau$ is the relaxation time, $\tau_0$ is the characteristic relaxation time, K is the anisotropy barrier, v is the volume, $k_B$ is the Boltzmann constant and $T_0$ is the Vogel-Fulcher temperature, which represents the temperature at which all spins freeze.[87-89] $Kv/k_B$ is often referred to as the activation energy $E_a$ (in K). Fitting the frequency-dependent $T_f$ values of UNP, Fig. 6C, yielded an $R^2$ value of 0.99932, indicating exceptional agreement between the model and experimental data.

The activation energy, $E_a = (174.71 \pm 1.26)$ K, corresponds to a low anisotropy barrier of $K = (1.12 \pm 0.01)$ kJ m$^{-3}$, which is characteristic of glassy dynamics. Such dynamics arise from competing interactions that weaken the anisotropy barrier. For a canonical spin-glass, the ratio $E_a/T_0 < 1$, while for a clustered-spin-glass, $E_a/T_0 > 1$. Here, with $T_0 = 23.28$ K, a ratio of $E_a/T_0 = 7.505$ strongly suggests a clustered-spin-glass behaviour, characterised by magnetic clusters with complex short-range interactions.[82, 84, 90, 91] The derived relaxation time is notably large, $\tau_0 = (7.34 \pm 0.31)$E-7 s, and reflects the collective behaviour of spins within clusters. This behaviour is shaped by surface effects, which introduce additional competing spin interactions.[82-84, 91-93] The absence of this transition in HNP suggests that the $\gamma$-Fe$_2$O$_3$ shell enhances surface anisotropy, suppressing the competing interactions necessary for the spin-glass formation.

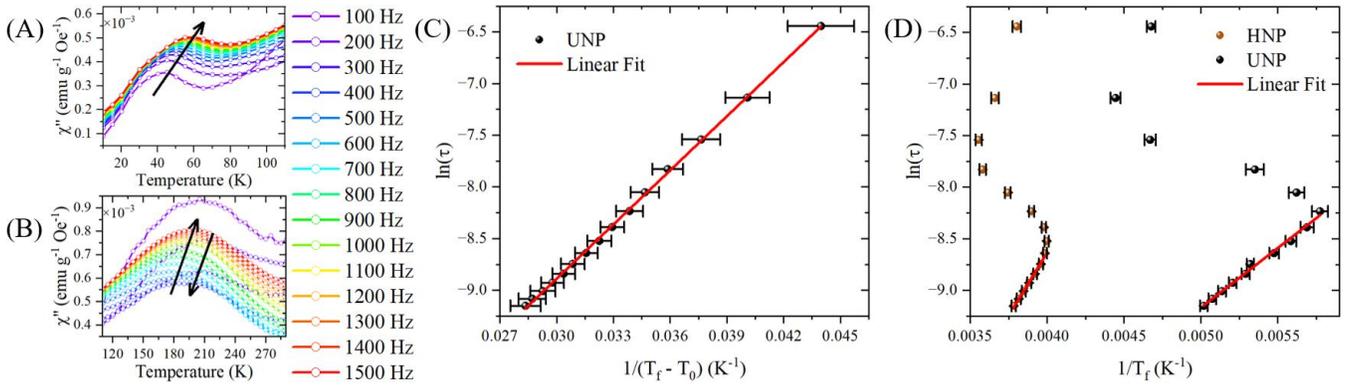

Figure 6. (A) and (B) Frequency dependence of the clustered-spin-glass and superparamagnetic transitions in UNP, respectively, with arrows indicating the shifting of peak positions with frequency. (C) Fitting the frequency dependence of the clustered-spin-glass transition with the Vogel-Fulcher Law (red). (D) Fitting the frequency dependence of the superparamagnetic transition with the Néel-Arrhenius Law (red).

At higher temperature (Fig. 6B), $T_f$ positions show an initial decrease followed by a gradual increase with respect to frequency. This behaviour reflects two distinct relaxation processes, each dominating different frequency regimes. Brownian relaxation, involving the physical rotation or translation of the entire nanoparticle, dominates at lower frequencies and in less viscous media. Meanwhile, Néel relaxation, the rotation of the magnetic moment within the nanoparticle itself, prevails at higher frequencies. In Fig. 6D, we clearly see the transition frequency from Brownian to Néel relaxation dynamics is at 600 Hz for the UNP, and 800 Hz for the HNP. This shift arises from the difference in surface states. Surface water and FeOOH increase the viscosity of the medium surrounding the UNP, preventing Brownian relaxation from dominating at higher frequencies. In HNP, this viscous layer is removed, so Brownian motion dominates to a higher frequency. Similar transition values have been published by S. Ota and Y. Takemura [94] and J. Dieckhoff et. al. [95], but are highly dependent on the nanoparticle size and composition. To analyze the superparamagnetic transition, the Néel-Arrhenius relation [96] can be implemented:

$$\tau = \tau_0 \exp\left(\frac{Kv}{k_B T_f}\right) \quad (3)$$

The difference with this model is the lack of the $T_0$ parameter, which accounts for collective spin interactions. Due to the transition from Brownian to Néel relaxation, we fit the Néel-Arrhenius model to the higher frequency tail-end of the data (600 Hz-1500 Hz and 900 Hz-1500 Hz for UNP and HNP respectively), as in Fig. 6D. The high $R^2$ values of 0.99644 and 0.99638 for UNP and HNP respectively, validate the linear relationship between $\ln(\tau)$ and $1/T_f$, and the use of the Néel-Arrhenius model. The activation energy of UNP, $E_a = (1155.84 \pm 34.59)$ K, corresponds to a magnetic anisotropy barrier of $K = (7.44 \pm 0.22)$ kJ m$^{-3}$, characteristic of superparamagnetic dynamics. In HNP, we observe a significant increase of $E_a = (2439.22 \pm 93.07)$ K, with $K = (15.69 \pm 0.59)$ kJ m$^{-3}$. This substantial increase of 111% in anisotropy energy indicates enhanced stability against thermal fluctuations, and explains the increased blocking temperature and vanishing of the clustered-spin-glass transition observed in UNP. The derived relaxation times for UNP and HNP are $\tau_0 = (3.23 \pm 0.60)$E-7 s and $\tau_0 = (1.03 \pm 0.37)$E-8 s, respectively, indicating a 97% faster relaxation time in HNP. The significant decrease in $\tau_0$ coupled with the increase in K upon heat-treatment is key to understanding the difference in SAR results.

The increase in K and decrease in $\tau_0$ observed in HNP significantly enhances the SAR, despite the reduction in saturation magnetization, which would typically lower SAR. Interestingly, despite the increase in K, we also see a reduction in the magnetic coercivity of HNP, which becomes more prominent at lower temperature. This may be attributed to suppressed spin-glass freezing [97, 98] or the strain-induced spin disorder from the $\gamma$-Fe$_2$O$_3$ shell.[80, 99] For example, L. Qian et. al.[99] show that the annealing temperature of Fe$_3$O$_4$ nanoparticles determines the coercivity, which decreases as the annealing temperature increases. As previously mentioned, for our 16 nm diameter particles, Néel relaxation is the dominant mechanism for heat generation at higher frequencies. The higher anisotropy in HNP creates an energy barrier for magnetic moment reversal that aligns more effectively with typical AC magnetic field frequencies used in hyperthermia, maximising energy dissipation. Similar findings have been published by R. Das et. al. [100] where Fe$^{2+}$ ions are replaced by Co$^{2+}$ ions in Fe$_3$O$_4$ to increase magnetic anisotropy, resulting in an increased SAR value. R. Das et. al. [101, 102] also show that forming Fe$_3$O$_4$ nanorods with higher surface anisotropy than spherical nanoparticles also increases the SAR. Here, we report that just changing the surface states and cell strain of Fe$_3$O$_4$ is sufficient enough to change the magnetic anisotropy. Furthermore, the large decrease in relaxation time indicates that spin flip events occur more frequently, leading to an increase in the number of relaxation processes. This higher frequency of relaxation enhances the system's efficiency, thereby optimizing the SAR. It is important to note that there is a cutoff where increasing the anisotropy too high will eventually inhibit the ability of spins to align with the AC field and lower the SAR. [100, 103]

CONCLUSIONS

In this study we have shown that heating treating Fe$_3$O$_4$ nanoparticles removes the adsorbed surface water and FeOOH, dehydrating them to form a shell of $\gamma$-Fe$_2$O$_3$ around the nanoparticles. The altered surface state results in a significant increase in the AC induction rate and consequently the SAR by 140% in the 0.315 g sample, which we accredit to the 111% increase in magnetic anisotropy and 97% faster relaxation time of the superparamagnetic nanoparticles. The increased anisotropy also suppresses the competing interactions required for the low-temperature spin-glass transition.

MATERIALS AND METHODS

**Sample preparation.** 25g of 99.9% purity, 8 nm radius Iron Oxide ($Fe_3O_4$) nanoparticles were purchased from US Research Nanomaterials, Inc. These particles were manufactured by laser synthesis and appear dark brown and black in color. 12.5g of power was set aside and left untreated (UNP) (inset of Figure 1B). The other 12.5g was placed in a glass dish, loosely covered with Al foil and placed into a Memmert vacuum oven. The environment was evacuated to vacuum before a Nitrogen ($N_2$) gas atmosphere was established. Nanoparticles were then annealed for 4 hours at 473K before being removed from the oven as a reddish-brown color (HNP) (inset of Figure 1B).

**Synchrotron powder diffraction.** High intensity synchrotron powder diffraction patterns of the samples were measured at the Australian synchrotron, using monochromatic radiation with $\lambda = 0.688733$ Å and energy E = 18 keV. The scattering angle (2θ) ranged between 2° and 82° with step size of 0.00375°. Data collected were analyzed in GSAS-II [104], through Rietveld refinement. Images of the unit cell were produced in VESTA [105].

**Thermogravimetric analysis.** Thermogravimetric analysis (TGA) was performed on UNP by heating at a constant rate of 5 °C min$^{-1}$ from room temperature to 400 °C under an inert argon atmosphere, followed by isothermal holding at 400 °C for 4 hours, to replicate prior annealing conditions and assess the structural stability of the material.

**Transmission electron microscopy.** Transmission electron microscopy (TEM) was used to reveal the crystallinity and size distribution of the nanoparticles. Experiments were conducted on the JEOL JEM-F200 at the Electron Microscopy Centre (EMC) at the University of Wollongong (UOW). Brightfield TEM images were analyzed using the Gatan Microscopy software suite (Version 2.32.888.0).

**Scanning electron microscopy and energy dispersive spectroscopy.** Scanning electron microscopy (SEM), was used to reveal the morphology of nanoparticles and to perform energy dispersive spectroscopy (EDS) mapping. The Aztec software suite (version 4.2) was used to analyze EDS images. Experiments were conducted on the JEOL JEM-F200 at the Electron Microscopy Centre (EMC) at UOW.

**X-ray photoelectron spectroscopy.** The X-ray photoelectron spectra (XPS) were collected on a NEXSA X-ray photoelectron spectrometer (Thermo Fisher Scientific) at UOW, with a monochromatized Al Kα excitation source (E = 1486.6 eV) to obtain the surface valence states. An ion gun was used to etch the nanoparticle surface for 50 s to reveal the core structure for analysis. A charge correction was performed by referencing the C 1s peak of carbon at 284.8 eV, to correct binding energy shifts caused by surface charging during measurement

**X-ray absorption spectroscopy.** Measurements were taken on the soft x-ray spectroscopy (SXR), x-ray absorption spectroscopy (XAS) at the Australian synchrotron. The sample was wrapped in an Fe foil standard. No standards were used for O. Energy scans were taken from 660 eV – 740 eV for the Fe L edge and from 520 eV – 560 eV for the O K edge.

**Vibrating sample magnetometry.** Vibrating sample magnetometry (VSM) was performed on a Physical Property Measurement System (PPMS) (Quantum Design, USA) to measure the magnetic properties of the $Fe_3O_4$ nanoparticles. The samples were mounted in a low-background, plastic sample holder to conduct the measurements. To measure hysteresis loops, samples were scanned between -5000 Oe and 5000 Oe. Samples were measured from 5 K to 350 K in 50 K intervals to determine temperature dependence. To detect the superparamagnetic transition and determine the blocking temperature, field-cooled and zero-field cooled temperature dependent data were collected between 2 K and 350 K.

**Time-of-flight inelastic neutron scattering.** Time-of-flight inelastic-neutron-scattering (TOF-INS) was used to investigate the quasielastic neutron scattering (QENS) signal in both $Fe_3O_4$ nanoparticle samples. TOF-INS data were collected on the Pelican instrument at the Australian Nuclear Science and Technology Organization (ANSTO). The Pelican instrument was operated with a neutron wavelength of $\lambda$ =4.69 Å, which afforded an energy resolution at the elastic line of 0.13 meV. 12.5 g of each sample were placed in separate annular Al cans with an overall neutron path length of 4 mm, to reduce multiple scattering, and data were collected at 200 K and 300 K for 5 hours. An empty can sample was collected for background subtraction and detector efficiencies were normalized with a vanadium can. The Large Array Manipulation Program (LAMP) was used for data analysis to extrapolate the scattering function $S(Q,\omega)$ and generalized density of states (DOS).

**Induction efficacy.** To investigate AC induction properties, a custom apparatus was designed to evaluate the heating efficiency. A power source (GW INSTEK PSM 3004) was used to power the coil with a current of 2.8 A and operated at a frequency of 85 kHz. This produced a magnetic field with an amplitude of ~ 4 Oe at the center of the coil. To minimize the convection heat transfer from the copper coils to the sample, cool water flowed through the void in the induction coil, maintaining ambient temperatures. To record the sample temperature, a FLIR MSX Integrated Optical System was used to capture thermal images in real time. We tested two different sample masses, 0.315 g and 5 g, to examine how sample size affects heating performance and to explore potential scaling effects.

**AC Magnetometry.** AC magnetometry was performed on a Physical Property Measurement System (PPMS) (Quantum Design, USA). 0.022 g of UNP and 0.018 g of HNP were prepared in bottom of a small capsule with a solid solution of eicosane placed on top to allow samples to move under Brownian motion but not escape the capsule during the measurement. Scans were taken from 10 K-300 K, in increments of 5 K, over a frequency range of 100 Hz-1500 Hz, in increments of 100 Hz, with a 5 Oe oscillating AC field. To measure the nature of transitions, the same scans were also conducted in an external DC field of 100 Oe and 1000 Oe.

**Supporting Information**
Scattering function map of water at room temperature, QENS analysis of UNP and HNP at 200 K and 300 K, imaginary susceptibility in an external DC field.

**Corresponding Authors**

* **Kyle A. Portwin** – Institute for Superconducting and Electronic Materials, Faculty of Engineering and Information Science, University of Wollongong, NSW 2500, Australia.
Email: kyleportwin@gmail.com

† **Zhenxiang Cheng** – Institute for Superconducting and Electronic Materials, Faculty of Engineering and Information Science, University of Wollongong, NSW 2500, Australia.
Email: cheng@uow.edu.au

**Authors**

**Pablo Galaviz** – Australian Centre for Neutron Scattering, Australian Nuclear Science and Technology Organization, Lucas Heights, New South Wales 2234, Australia

**Xiaoning Li** – Institute for Superconducting and Electronic Materials, Faculty of Engineering and Information Science, University of Wollongong, NSW 2500, Australia

**Chongyang Hao** – Institute for Superconducting and Electronic Materials, Faculty of Engineering and Information Science, University of Wollongong, NSW 2500, Australia

**Lachlan A. Smillie** – Institute for Superconducting and Electronic Materials, Faculty of Engineering and Information Science, University of Wollongong, NSW 2500, Australia

**Mengyun Yu** – Institute for Superconducting and Electronic Materials, Faculty of Engineering and Information Science, University of Wollongong, NSW 2500, Australia

**Caleb Stamper** – Institute for Superconducting and Electronic Materials, Faculty of Engineering and Information Science, University of Wollongong, NSW 2500, Australia

**Richard Mole** – Australian Centre for Neutron Scattering, Australian Nuclear Science and Technology Organization, Lucas Heights, New South Wales 2234, Australia

**Dehong Yu** – Australian Centre for Neutron Scattering, Australian Nuclear Science and Technology Organization, Lucas Heights, New South Wales 2234, Australia

**David L. Cortie** – Australian Centre for Neutron Scattering, Australian Nuclear Science and Technology Organization, Lucas Heights, New South Wales 2234, Australia

**Kirrily C. Rule** – Australian Centre for Neutron Scattering, Australian Nuclear Science and Technology Organization, Lucas Heights, New South Wales 2234, Australia

**Author Contributions**

The manuscript was written through contributions of all authors.

**Funding Sources**

This project was funded as a part of the Australian Research Council (ARC) Discovery Project (DP) DP210101436.

**Notes**

The authors declare no competing financial interest.

## ACKNOWLEDGMENT

Neutron beam and scientific computing time was awarded at the Australian Centre for Neutron Scattering (ANSTO) under proposal P15798 and synchrotron beamtime was awarded at the Australian Synchrotron (ANSTO). Electron microscopy was supported by the Electron Microscopy Centre (EMC) at the University of Wollongong. KAP is supported by the postgraduate research award (PGRA) provided by the Australian Institute of Nuclear Science and Engineering (AINSE), the Australian Government Research Training Program (AG-RTP), and KAP also received an honours scholarship from AINSE.

## ABBREVIATIONS

UNP, untreated nanoparticles; HNP, heat-treated nanoparticles; SAR, specific absorption rate; FWHM, full-width-at-half-maximum; PDF, pair distribution function; TGA, thermogravimetric analysis; XPS, x-ray photoelectron spectroscopy; XAS, x-ray absorption spectroscopy; TEM, transmission electron microscopy; SEM, scanning electron microscopy; EDS, electron dispersive spectroscopy; TOF-INS, time-of-flight inelastic neutron spectroscopy; DOS, density of states; QENS, quasielastic neutron scattering; PPMS, physical property measurement system; DC, direct current; AC, alternating current; EISF, elastic incoherent structure factor.

# Tuning the Surface States of Fe₃O₄ Nanoparticles for Enhanced Magnetic Anisotropy and Induction Efficacy

Kyle A. Portwin,[1,a)] Pablo Galaviz,[2] Xiaoning Li,[1] Chongyan Hao,[1] Lachlan A. Smillie,[1] Mengyun You,[1] Caleb Stamper,[1] Richard Mole,[2] Dehong Yu,[2] Kirrily C. Rule,[1,2] David L. Cortie,[2] and Zhenxiang Cheng[1,b)]

Affiliations

[1] Institute for Superconducting and Electronic Materials, Faculty of Engineering and Information Science, University of Wollongong, NSW 2500, Australia.

[2] Australian Centre for Neutron Scattering, Australian Nuclear Science and Technology Organization, Lucas Heights, New South Wales 2234, Australia

Authors to whom correspondence should be addressed: [a)]kyleportwin@gmail.com  [b)]cheng@uow.edu.au

**Neutron Inelastic Rietveld Refinements**

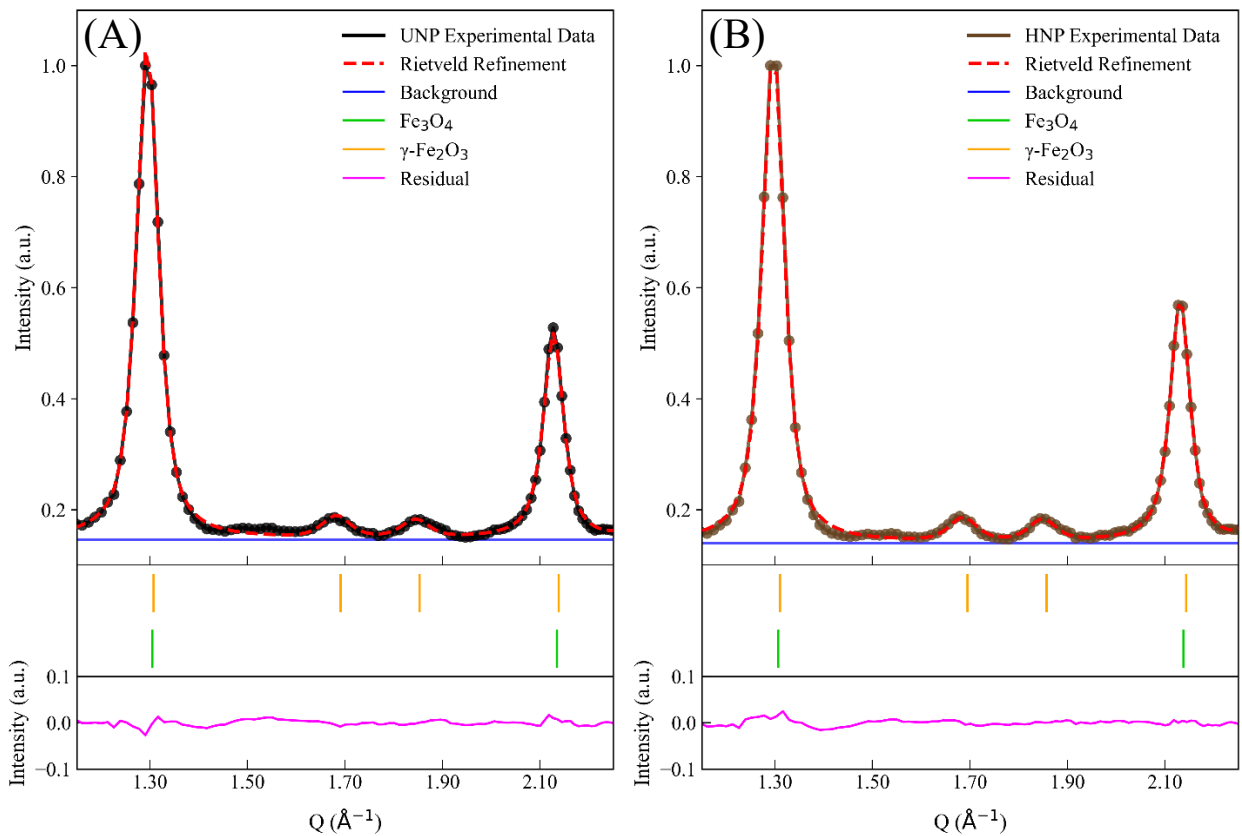

*Figure S1. Inelastic neutron scattering with Rietveld refinements for (A) UNP and (B) HNP. Below the data and refinements are the tick marks for the $Fe_3O_4$,[106] and $\gamma\text{-}Fe_2O_3$,[107] phases, followed by the residual plot of refinements subtracted from the experimental data. Experimental data are normalized by the most intense reflection to 1.*

Rietveld refinements of the elastic signal from time-of-flight inelastic neutron scattering, functionally equivalent to neutron powder diffraction, reveals clear structural differences between UNP and HNP. Unlike X-ray/synchrotron powder diffraction, which is highly surface sensitive and primarily probes the nanoparticle shell, neutron diffraction offers a bulk-sensitive perspective, allowing for simultaneous characterization of both surface and core phases. We perform Rietveld refinements of UNP and HNP, Fig. S1(A) and (B), respectively, at 200 K to reduce the background signal from water. The refinements were conducted with key structural parameters constrained to isolate meaningful differences in phase composition and lattice metrics. For both refinements, the particle size was fixed at 16 nm and microstrain held constant at 1000, while instrumental and sample-dependent parameters were kept uniform to eliminate external sources of variability. The only free parameters in these fits were the phase fractions and the cubic lattice parameters of $Fe_3O_4$ and $\gamma\text{-}Fe_3O_4$.

This refinement is challenging due to the substantial peak overlap between $Fe_3O_4$ and $\gamma$-$Fe_2O_3$, two Bragg reflections are shared by both phases, while only two reflections are unique to $\gamma$-$Fe_2O_3$. Despite this, the refinements robustly resolve the dominant phase as $Fe_3O_4$ in both samples, as evidenced by the high weight fractions of 87.6(4)% and 86.2(5)% for UNP and HNP, respectively, presented in Table S1. This majority $Fe_3O_4$ content is also reflected in the refined lattice parameters, which remain closer to the expected bulk value for $Fe_3O_4$ (8.396 Å) than for $\gamma$-$Fe_2O_3$ (8.33–8.35 Å). In the HNP sample, the increased $\gamma$-$Fe_2O_3$ content is indicated by both a higher weight fraction and a modest decrease in the refined lattice parameters, consistent with the incorporation of a greater proportion of the oxidized spinel phase. These results underscore the sensitivity of neutron diffraction to subtle structural changes across the nanoparticle ensemble and confirm that not only are the nanoparticles a majority $Fe_3O_4$ core with $\gamma$-$Fe_2O_3$ shell, heat treatment induces partial oxidation of $Fe_3O_4$ to $\gamma$-$Fe_2O_3$, leading to measurable shifts in both phase composition and lattice geometry.

**Table S1. Rietveld refinement parameters**

|  | UNP | HNP |
|---|---|---|
| wR | 2.54% | 2.18% |
| R | 2.04% | 1.99% |
| $Fe_3O_4$ weight fraction | 87.6(4)% | 86.2(5)% |
| $Fe_3O_4$ lattice parameter (a = b = c) | 8.3918(23) Å | 8.3794(16) Å |
| $\gamma$-$Fe_2O_3$ weight fraction | 12.4(4)% | 13.8(5)% |
| $\gamma$-$Fe_2O_3$ lattice parameter (a = b = c) | 8.375(14) Å | 8.357(11) Å |

**$S(Q,\omega)$ Map.** Fig. S2 presents the QENS signal for water at room temperature to compare with the QENS of the $Fe_3O_4$ nanoparticle samples. There is an absence of any Bragg peaks in the $S(Q)$ plot, due to the liquid dynamics of the water. Notably, the QENS signal continues to grow as Q is increased, whereas in the nanoparticles, the QENS begins to flatten, suggesting there is some additional signal from other hydrogen and/or hydroxyl groups such as FeOOH.

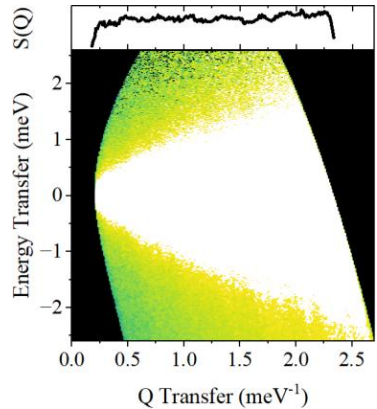

*Figure S2. $S(Q,\omega)$ map for $H_2O$ at room temperature.*

**QENS Analysis.** It is standard in QENS analysis to plot the FWHM (meV) against $Q^2$ (Å$^{-2}$) and the Elastic Incoherent Structure Factor (EISF) (a.u.) against Q (Å$^{-1}$), Fig. S3A and S3B display this analysis, respectively. Firstly, it is important to note that there are false signals in the QENS where there are Bragg peaks and acoustic modes. In $Fe_3O_4$/$\gamma$-$Fe_2O_3$, Bragg peaks appear at 1.3 Å$^{-1}$, 1.7 Å$^{-1}$, 1.85 Å$^{-1}$ and 2.1 Å$^{-1}$, which is why the corresponding results here don't follow the natural trend. Results are also skewed at low Q in the 200 K samples, including a peak at ~0.3 meV. This is a spurious feature from the cryofurnace on Pelican, so should be ignored.

At 300 K, the FWHM appears to increase linearly with respect to $Q^2$ at low values. In UNP the FHWM continues to grow, whereas the HNP begins to flatten out. The linear increase is characteristic of liquid-like diffusion dynamics. When the FWHM begins to flatten, this is a sign that there is rotational or confined diffusion, as opposed to long-range diffusion. At 200 K the signals from both samples are quite contaminated by the spurious feature and Bragg peaks, however it does appear flatter than the room temperature samples, which is expected.

The EISF data shows a much clearer picture than FWHM data. A flat EISF corresponds to localized motion or rotational diffusion, while an exponential decay indicates the presence of long-range translational diffusion. There is a clear difference here between the UNP at 300 K, which has an exponential decay, until reaching the Bragg peak at 1.3 Å$^{-1}$, and all other samples/temperatures which show a flat EISF. Interestingly, we can see that the exponential decay in UNP at 300 K does not continue after the Bragg peak, which tells us at short lengths, rotational or localized diffusion begins to dominate.

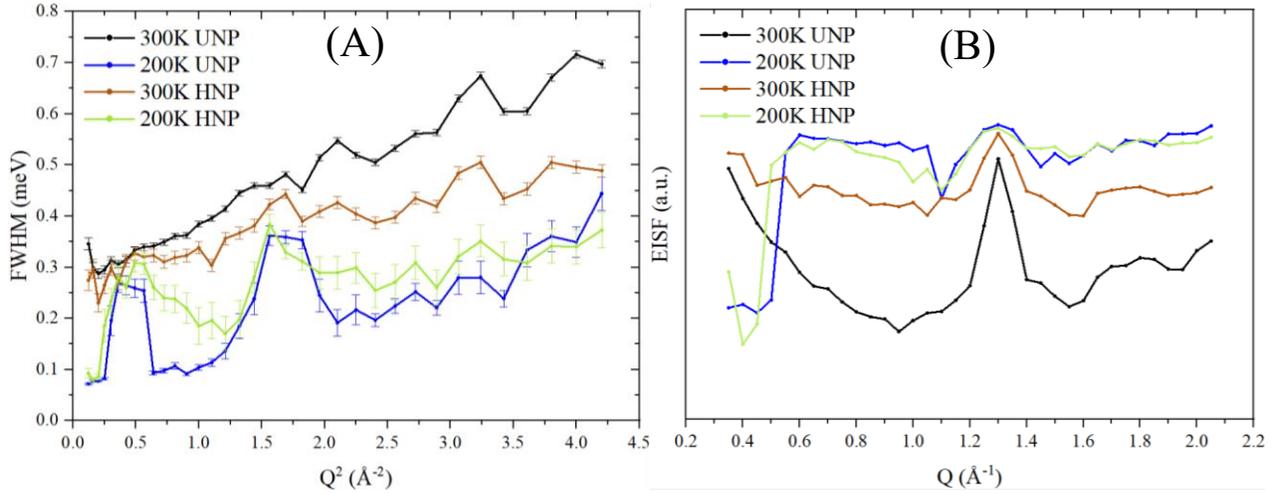

Figure S3. QENS analysis, (A) plots the FWHM against $Q^2$ and (B) shows the EISF against Q.

**Imaginary AC susceptibility in an external DC field.** Fig. S4 presents the temperature dependent $\chi''$ of UNP and HNP at different frequencies and applied external DC fields. The low-temperature peak in UNP, which is evident across all frequencies, shifts slightly to lower temperatures when an external DC magnetic field is applied. This subtle shift suggests that the low-temperature peak corresponds to a clustered-spin-glass transition. Spin glasses are characterised by disordered and frustrated spin interactions, where clusters of spins freeze into a non-equilibrium state at low temperatures. When an external field is applied, the spins partially align with the field, reducing random spin fluctuations, so a lower temperature is required to freeze the spins into random orientations [82, 84]. In contrast, the high-temperature peak in $\chi''$, associated with the superparamagnetic relaxation of nanoparticles, shows a dramatic shift to higher temperatures with increasing external field strength. The super-paramagnetic phase is characterised by spontaneous spin-flipping events in the magnetic nanoparticles. When an external field is applied the spins align with the field, stabilising the energy barriers associated with magnetic anisotropy. This means a higher temperature is required to provide the spins enough thermal energy to overcome the increased anisotropy barriers. We can also clearly see here that the HNP have a higher magnetic anisotropy as they naturally have a higher blocking temperature [97, 108]. Both the low-temperature spin glass and high-temperature superparamagnetic peaks exhibit these field-induced shifts across the frequency range studied.

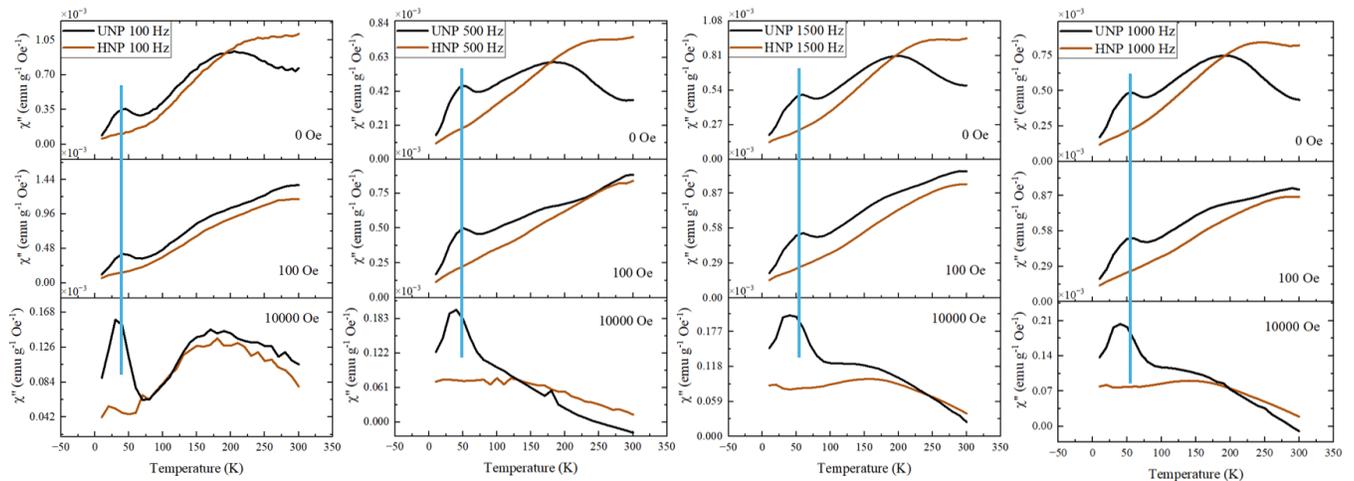

Figure S4. Imaginary susceptibility as a function of temperature at different frequencies and applied DC fields for UNP and HNP. The blue line highlights the shift in peak position with applied field.